\documentclass[aps,prl,twocolumn,superscriptaddress,10pt]{revtex4-2}

\usepackage{amsmath, amssymb}
\usepackage{mathrsfs}
\usepackage[dvipdfmx]{graphicx}
\usepackage{bm}
\usepackage{braket}
\usepackage{multirow}
\usepackage[colorlinks=true, citecolor=blue]{hyperref}
\usepackage{here}

\newcommand{\del}{\partial}
\newcommand{\diff}{\mathrm{d}}
\newcommand{\imag}{\mathrm{Im}\,}

\newcommand{\Hc}{\mathrm{H.c.}}

\newcommand{\imu}{\mathrm{i}}
\newcommand{\epn}{\mathrm{e}}

\newcommand{\ua}{\uparrow}
\newcommand{\da}{\downarrow}
\newcommand{\dg}{\dagger}
\newcommand{\la}{\langle}
\newcommand{\ra}{\rangle}

\newcommand{\sg}{\sigma}
\newcommand{\om}{\omega}
\newcommand{\gm}{\gamma}
\newcommand{\lam}{\lambda}
\newcommand{\ep}{\varepsilon}
\newcommand{\T}{\mathrm{T}}
\newcommand{\dvec}[1]{\hspace{-1mm}\stackrel{\leftrightarrow}{#1}\hspace{-1mm}}
\newcommand{\nt}{\notag \\}

\newcommand{\mrm}[1]{\mathrm{#1}}
\newcommand{\mcal}[1]{\mathcal{#1}}
\newcommand{\mscr}[1]{\mathscr{#1}}

\begin{document}

\title{
Quantification of electronic asymmetry: chirality and axiality in solids
}

\author{Tatsuya Miki}
\affiliation{Department of Physics, Saitama University, Sakura, Saitama 338-8570, Japan}
\author{Hiroaki Ikeda}
\affiliation{Department of Physics, Ritsumeikan University, Kusatsu, Shiga 525-8577, Japan}
\author{Michi-To Suzuki}
\affiliation{Department of Materials Science, Graduate School of Engineering, Osaka Metropolitan University, Sakai, Osaka 599-8531, Japan}
\affiliation{Center for Spintronics Research Network, Graduate School of Engineering Science, Osaka University, Toyonaka, Osaka 560-8531, Japan}
\author{Shintaro Hoshino}
\affiliation{Department of Physics, Saitama University, Sakura, Saitama 338-8570, Japan}

\date{\today}

\begin{abstract}

Chiral and axial materials offer platforms for intriguing phenomena, such as cross-correlated responses and chirality-induced spin selectivity. 
However, quantifying the properties of such materials has generally been considered challenging. 
Here, we demonstrate that the spatial distribution of the electron chirality, represented by $\Psi^\dg \gamma^5 \Psi$ with the four-component Dirac field $\Psi$, characterizes the chirality and axiality of materials. 
Furthermore, we reveal that spin-derived electric polarization can serve as an effective indicator of material polarity.
We present quantitative evaluations of electron chirality distribution and spin-derived electric polarization based on first-principles calculations.
Additionally, we propose that electron chirality can be directly observed via circular dichroism in photoemission spectroscopy, which measures the difference between right- and left-handed circularly polarized light. 
Electron chirality and spin-derived electric polarization provide a new framework for quantifying chirality, axiality, and polarity in asymmetric materials, paving the way for the exploration of novel functional materials.

\end{abstract}

\maketitle

Materials with chiral or axial structures offer new possibilities for various responses that reflect their inherent asymmetry. Notable examples include cross-correlated response, chiral magnetism \cite{Tokura18, Tokura21}, chirality-induced spin selectivity (CISS) \cite{Gohler11, Naaman15, Naaman20, Inui20, Bloom24}, and circular dichroism \cite{Brinkman24}. Furthermore, the concept of chirality extends beyond condensed matter physics into fields such as biology, chemistry, and particle physics.

In 1964, Lipkin introduced a conserved quantity in vacuum for electromagnetism, $Z = \bm E \cdot (\nabla \times \bm E) + \bm B \cdot (\nabla \times \bm B)$, which he referred to as ``zilch'' due to its unclear physical significance at that time \cite{Lipkin64}. 
However, it was later discovered that Lipkin's zilch is proportional to the asymmetry factor of circular polarization in light absorption, thus becoming known as optical chirality \cite{Tang10}.

In chemistry, molecular chirality---or handedness---is crucial because different enantiomers often exhibit distinct biological behaviors. Understanding and controlling this chirality is of critical importance.
To achieve this, it is essential to quantify the degree of chirality in materials, yet this has traditionally been a challenging task. For example, while one might attempt to quantify molecular chirality based on atomic positions \cite{Harris97}, this definition is not uniquely determined \cite{Harris99}.

In this paper, we focus on the chirality of electrons, as it appears in the Dirac field of relativistic quantum theory. 
In condensed matter physics, attention is typically given to the charge and spin degrees of freedom of electrons, but chirality is also a fundamental degree of freedom for electrons.
Electron chirality is even under time reversal and locally breaks both inversion and mirror symmetries, making it a suitable measure to describe the chirality of materials.
By focusing on electron chirality rather than spin, we aim to explore quantitative expressions for asymmetric materials,
suggesting the possibility of uncovering new insights that have been overlooked in non-magnetic materials. 
This study presents a first-principles evaluation showing that the distribution of electron chirality serves as a key factor in quantifying the degree of chirality in asymmetric materials. This is the central theme of this paper.

Eelectron chirality is defined as the time-component of the four-component pseudovector in terms of Lorentz transformation, which is represented by the gamma matrix $\gm^5$ \cite{Berestetskii_book, Sakurai_book}.
For the introduction of physical quantities, we here employ the Weyl representation of a four-component Dirac field $\Psi(\bm r) = (\psi_{\mrm R}(\bm r), \psi_{\mrm L}(\bm r))^\T$, where $\psi_{\mrm R(\mrm L)}(\bm r)$ denotes a two-component field operator for the right(left)-handed electron at the position $\bm r$.
In this representation, the chirality density $\tau^Z(\bm r) = \Psi^\dg(\bm r) \gm^5 \Psi(\bm r)$ is given as the difference between electrons with right-handed (R) and left-handed (L) helicities \cite{Berestetskii_book, Sakurai_book, Fukuda16, Senami19, Senami19, Hoshino24}:
\begin{align}
    \tau^Z(\bm r) &= \psi^\dg_{\mrm R} (\bm r) \psi_{\mrm R}(\bm r) - \psi^\dg_{\mrm L}(\bm r)\psi_{\mrm L}(\bm r), \label{eq:chirality_dirac}
\end{align}
which is regarded as a material version of ``zilch''.
$\tau^Z(\bm r)$ is defined at each spatiotemporal point and characterizes the chiral state of materials microscopically.
From the perspective of transformation properties under spatial inversion (SI), chirality behaves as pseudoscalar (SI odd), which is directly confirmed by the transformation law for the Weyl spinor $\psi_{\mrm R(\mrm L)} \to \psi_{\mrm L(\mrm R)}$ by SI.

The electron chirality at the position of the nucleus is related to the parity violation energy difference between enantiomers \cite{Vester59, Ulbricht62, Andrea99, Laerdahl00, Senami19, Kuroda22}.
The spatial distribution of electron chirality has been calculated in quantum chemistry \cite{Bast11, Senami19, Kuroda22}, where the spatial gradient of the electron chirality density creates additional spin torque \cite{Tachibana12, Hara12, Fukuda16}.

\begin{figure}[tb]
    \centering
    \includegraphics[width=8.5cm]{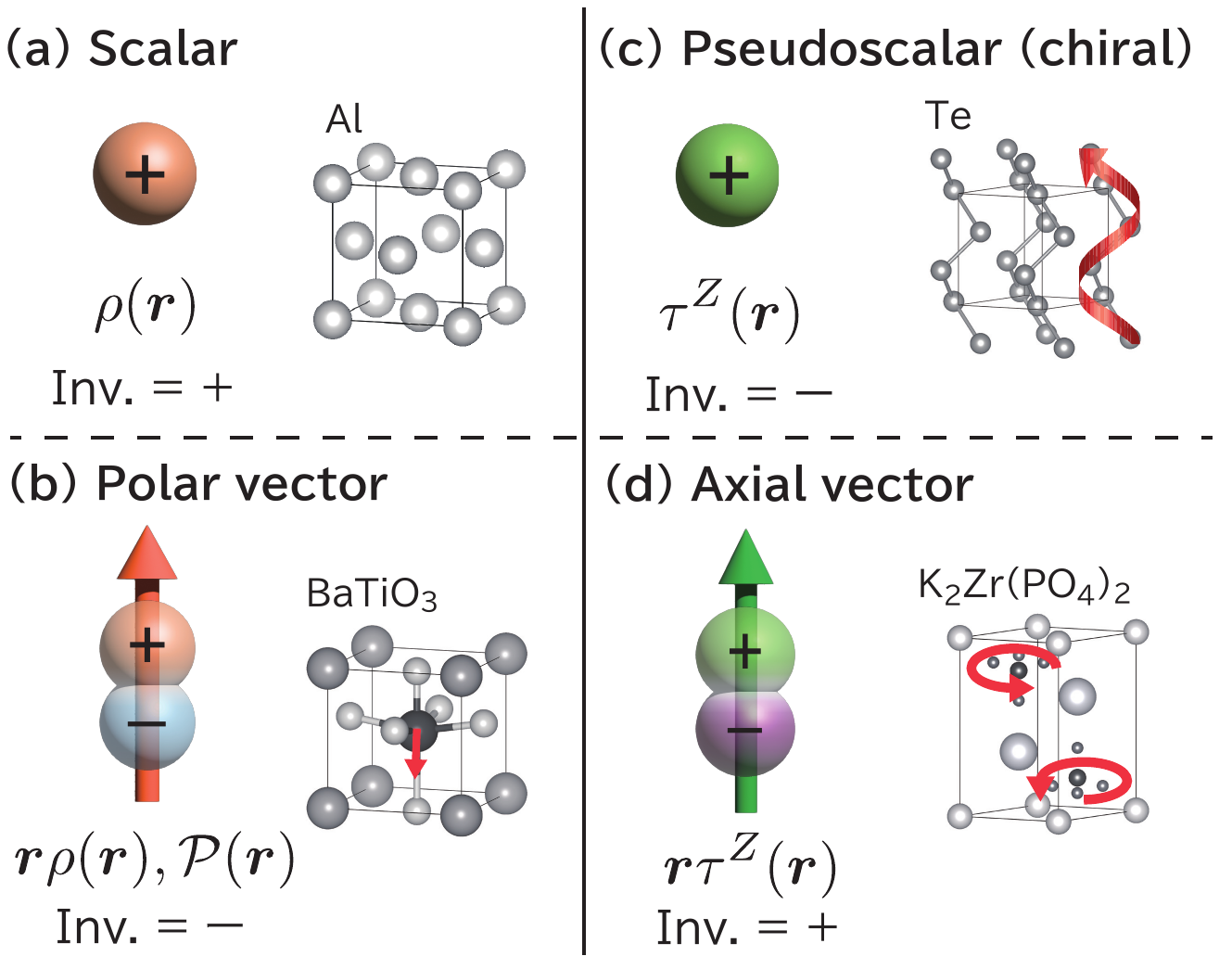}
    \caption{
    Quantification of asymmetric crystals.
    (Left) The charged (ionic) state and polar crystal are characterized by the charge density $\rho(\bm r)$ distribution.
    (Right) By contrast, the chiral and (ferro-)axial crystals are quantified based on the electron chirality $\gm^5$, i.e., $\tau^Z(\bm r)$ defined in Eq.~\eqref{eq:chirality_dirac}, which measures the difference between right- and left-handed particles.
    The symbol ``Inv.$=+(-)$'' indicates that each physical quantity is even (odd) under SI.
    For completeness, we show Al crystal as an example of a highly symmetric simple substance in (a).
    }
    \label{fig:concept}
\end{figure}

In the following, we demonstrate that the distribution of electron chirality in Eq.~\eqref{eq:chirality_dirac} serves as a microscopic quantity for quantitatively characterizing the degree of symmetry breaking in chiral \cite{Rao19, Sakano20, Inui20, Brinkman24} and (ferro-)axial materials \cite{Zhang15, Hlinka16, Jin20, Hayashida20, Hayashida21, Hayami22, Hayashida23, Yamagishi23}, offering a new perspective on asymmetric crystals.
As provided below, the physical quantities characterizing chirality and axiality can be defined based on the transformation properties of the Dirac field under SI.

The main findings of this paper are summarized in Fig.~\ref{fig:concept}.
In addition to the electron chirality, i.e. pseudoscalar (SI odd), we can introduce the quantity with scalar (SI even), polar vector (SI odd), and axial vector (SI even).
Whereas the electron chirality in Eq.~\eqref{eq:chirality_dirac} is defined as the difference between the left- and right-handed electron densities, the scalar quantity is obtained by their summation $\rho = \psi_{\mrm R}^\dg \psi_{\mrm R} + \psi_{\mrm L}^\dg \psi_{\mrm L}$, which is nothing but the electron charge density [Fig.~\ref{fig:concept} (a)].
The integral value $N = \int\diff\bm r \la\rho(\bm r)\ra$ represents the total charge or the number of electrons.
On the other hand, the polar vector is obtained by the product between the position vector $\bm r$ and scalar [(SI odd) = (SI odd) $\times$ (SI even)], namely $\bm r\rho$.
If the material has a finite integral value of $\bm P = \int \diff \bm r \la \bm r \rho(\bm r) \ra$, then the material is said to have ``polarity'' or the electric polarization [the arrow in Fig.~\ref{fig:concept} (b)]
\footnote{
When we consider the bulk crystals, it is well known that $\bm r \rho(\bm r)$ in the electric polarization $\bm P$ is ill-defined. 
This issue has been addressed through the modern theory of polarization involving Berry curvature \cite{Resta92, King-Smith93, Vanderbilt93, Resta94}.
}.
$N$ is even under SI, while $\bm P$ is odd.

In a similar manner to the consideration of the polar vector, we can define an axial vector, which characterizes the axial materials, derived from the pseudoscalar, i.e. electron chirality.
Namely, we can obtain the axial vector as $\bm r \tau^Z(\bm r)$ [(SI even) = (SI odd) $\times$ (SI odd)], where we refer to this axial vector as ``axiality''.
The expression of $\bm r\tau^Z(\bm r)$ indicates that the polarization of chirality specifies the axiality [the arrow in Fig.~\ref{fig:concept} (d)].
Just as the charge distribution reveals whether a material is non-polar or polar, the distribution of electron chirality makes it immediately apparent whether a material is chiral or axial.
The total chirality and axiality of material are evaluated by integrating their respective distributions, $C = \int \diff \bm r\, \la \tau^Z(\bm r) \ra$ (SI odd) and $\bm X = \int \diff \bm r\, \la \bm r \tau^Z(\bm r) \ra$ (SI even)
\footnote{
The axial vector $\bm X$ serves as order parameters representing the degree of axiality in a material, enabling quantitative analysis.
In the bulk crystal, a similar discussion to the Berry curvature formula for polarity $\bm P$ (see also, Ref.~\cite{Note1}) may be required for the axiality $\bm X = \int\diff\bm r\, \la\bm r \tau^Z(\bm r)\ra$. 
However, in systems with inversion symmetry such as $\mrm{K_2Zr(PO_4)_2}$, the origin can be chosen as the inversion center, and the unit cell can be defined as the Wigner-Seitz cell for defining $\bm X$.
We use this definition of $\bm X$ for computations in this paper.
For further details, refer to supplementary material \cite{supp}.
}
.

While we have considered the polar vector as $\bm r \rho(\bm r)$, we can also introduce the electric polarization which does not include the position vector $\bm r$.
Considering the transformation properties under SI, the three-component polar vector with SI odd can also be introduced by employing the three-component Pauli matrices \cite{Hoshino24}:
\begin{align}
    \bm {\mathcal P} (\bm r) &= -\imu \big[  \psi_{\mrm R}^\dg (\bm r)\bm \sg \psi_{\mrm L}(\bm r) - \psi_{\mrm L}^\dg (\bm r)\bm \sg \psi_{\mrm R}(\bm r)  \big].
    \label{eq:polar}
\end{align}
The same quantity is introduced from the Gordon decomposition \cite{Gordon28, Baym_book} or an equation of motion for electric current \cite{Hoshino24}.
For the polar crystals, we can utilize Eq.~\eqref{eq:polar} as a measure of polarity microscopically [Fig.~\ref{fig:concept} (b)].

We note that these characterizations can also be understood in the context of multipole \cite{Dubovik86, Dubovik90, Prosandeev06, Guo12, Hayami18, Hayami19, Kusunose20, Oiwa22, Hayami22, Hirose22, Kishine22, Hoshino23, Kusunose24, Inda24, Hayami24}.
The total chirality $C$ is regarded as an electric toroidal monopole \cite{Hoshino23}, and the total axiality $\bm X$ is regarded as an electric toroidal dipole.
These multipoles are unambiguously defined based on the electron chirality which is defined at each spatial point.
This is analogous to the standard electrostatics that the electric multipoles derive from the spatial distributions of charge density \cite{Jackson_book}.

\paragraph{Non-relativistic limit.}

In the above, we have introduced the electron chirality and the electric polarization in terms of the Dirac field described as four-component spinors, or right-/left-handed particle representation.
Now, let us focus on low-energy physics which is typically discussed in condensed matter physics.
We employ the non-relativistic limit (NRL) and consider the Schr\"{o}dinger field theory for two-component spinors $\psi = (\psi_\ua, \psi_\da)^\T$ ($\ua,\da$ is a spin degree of freedom).
The concrete expression is derived by $1/m$ expansion as performed in the derivation of relativistic correction for Hamiltonian such as Darwin term and spin-orbit coupling (SOC) \cite{Berestetskii_book, Bjorken_book, Frohlich93}.
We derive the expression for NRL of electron chirality and electric polarization in the supplementary material (SM) \cite{supp}.

The NRL of electron chirality is represented as the projection of the spin onto the direction of momentum, which is referred to as helicity:
\begin{align}
    \tau^Z(\bm r) = \frac{1}{2mc} \psi^\dg(\bm r) \dvec{\bm p} \cdot \bm \sg \psi(\bm r)
\end{align}
with $A \dvec{O} B = A O B + (O^*A) B$ \cite{Hoshino23, Hoshino24}.
The electric polarization in NRL consists of two parts expressed as $\bm{\mcal P} = \frac{1}{4mc} \bm \nabla (\psi^\dg \psi) + \bm{\mcal P}_S$ \cite{Wang06, Hoshino23, Hoshino24}. 
The first term represents a gradient of electronic density, where the Pauli matrices are not included.
Since the gradient $\bm \nabla (\psi^\dg \psi)$ does not contribute the integral $\int\diff\bm r \la \bm{\mcal P} \ra$, we concentrate on the second term $\bm{\mcal P}_S$ in the following.
We refer to this second term as the spin-derived electric polarization \cite{Hoshino23, Hoshino24}, whose expression is given by
\begin{align}
    \bm{\mcal P}_{S}(\bm r) = -\frac{1}{4mc} \psi^\dg(\bm r) \dvec{\bm p} \times \bm \sg \psi(\bm r).
\end{align}
We note that $\tau^Z$ and $\bm{\mcal P}_S$ can be expressed in a unified way using spin current tensor $j_{Sij} = \psi^\dg  \dvec{p_i} \sg^j \psi\, (i, j = x,y,z)$ as $\tau^Z(\bm r) = \frac{1}{2mc} \sum_i j_{Sii}(\bm r)$ and $\mcal P_{Si}(\bm r) = -\frac{1}{4mc} \sum_{jk} \epsilon_{ijk} j_{Sjk}(\bm r)$.
Namely, the symmetric part (trace) of the spin current tensor is $\tau^Z$, whereas the anti-symmetric part is $\bm{\mcal P}_S$ \cite{Hoshino23, Hoshino24}.

\paragraph{Evaluation of electron chirality.}

\begin{figure}[tb]
    \centering
    \includegraphics[width=8.5cm]{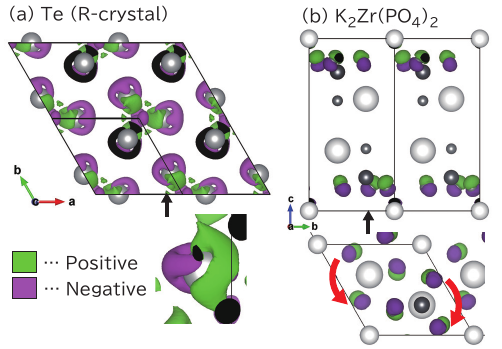}
    \caption{
    Spatial distributions of electron chirality $\tau^Z(\bm r)$ for (a) the chiral crystal Te and (b) the axial crystal K$_2$Zr(PO$_4$)$_2$.
    The black shadows indicate cross-sections of the unit cell.
    }
    \label{fig:space}
\end{figure}
Based on the above consideration, we turn to the quantitative evaluation of $\tau^Z$ for chiral crystal Te \cite{Sakano20} and axial crystal $\mrm{K_2Zr(PO_4)_2}$ \cite{Yamagishi23}.
$\tau^Z$ is evaluated using the first-principles band calculations \cite{supp}.

We begin with the spatial distribution of $\tau^Z$.
We expand the field operator by the Bloch function as $\psi = \sum_{n\bm k} \psi_{n\bm k} c_{n\bm k}$ and obtain the expectation value.
The electron chirality $\tau^Z$ for chiral crystal Te is shown in Fig.~\ref{fig:space} (a).
Focusing on a single Te atom [bottom figure in (a)], the electron chirality is distributed in wedge-shaped regions of positive and negative areas.

The distribution of electron chirality for axial crystal ${\rm K_2Zr(PO_4)_2}$ is shown in Fig.~\ref{fig:space} (b).
We can see a set of aligned chirality dipoles, which captures the characteristic of the axial crystal.
Accordingly, the distribution of electron chirality $\tau^Z$ distinguishes between chiral and axial materials.
We note that the axial structure in ${\rm K_2Zr(PO_4)_2}$ is formed due to a rotational distortion of oxygen atoms around the phosphorus atom, which is associated with mirror symmetry breaking from a non-axial configuration [indicated by the red arrows in Fig.~\ref{fig:concept} (d)].
As shown in the bottom figure of Fig.~\ref{fig:space} (b), the chirality dipoles are slightly canted around the rotational axis, which is indicated by the red arrows.

To clarify the behavior of the chirality dipole, we demonstrate the distribution of electron chirality in non-axial configuration ($P\bar 3m1$) \cite{Yamagishi23}, as shown in Fig.~\ref{fig:chirality} of SM \cite{supp}. 
In this configuration, the chirality dipole is already present and arranged in a circular structure around the $z$-axis, centered on each phosphorus atom. 
In the axial structure, the three oxygens surrounding each phosphorus atom rotate around the $z$-axis, inducing a field conjugate to the chirality dipole, similar to a rotational electric field ($\nabla \times \bm E$) \cite{Hoshino24},
along the $z$-axis.
This induced conjugate field results in a tilt of the chirality dipole along the $z$-axis.

In the following, we focus on the chiral crystal Te.
We consider the total values of $\tau^Z$ in materials, i.e., the unit cell integration, whose expression is given by
\begin{align}
    &C = \int_{\mrm{cell}} \diff \bm r \la\tau^Z(\bm r)\ra = \sum_{n\bm k} \tau_{n\bm k}^Z f_{n \bm k} \label{eq:chirality_bnd} 
\end{align}
where $\tau^Z_{n\bm k} = \frac{1}{2mc}\int\diff\bm r \psi_{n\bm k}^\ast \dvec{\bm p}\cdot\bm \sg \psi_{n\bm k}$ is a representation in the basis of Bloch function, and $f_{n\bm k}$ is an occupation function.
For evaluation of Eq.~\eqref{eq:chirality_bnd}, we employ Wannier interpolation to take a highly dense $\bm k$-mesh \cite{supp}.
We also compute the total axiality $\bm X = \int _{\rm cell} \diff\bm r \la \bm r \tau^Z \ra$ and the total polarity $\bar{\bm P} = \int_{\mrm{cell}} \diff\bm r \la \bm{\mcal P}_S(\bm r) \ra$, whose results are provided in the SM \cite{supp}.

\begin{figure}[tb]
    \centering
    \includegraphics[width=8.5cm]{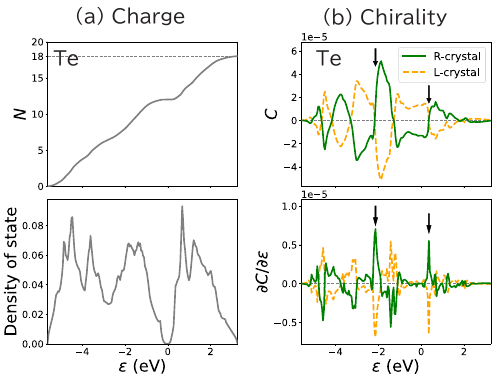}
    \caption{
    Energy dependences of (a) the electronic charge and (b) electron chirality for $\mrm{Te}$.
    The derivatives with respect to energies
    are also shown for both quantities.}
    \label{fig:integral}
\end{figure}
Figure~\ref{fig:integral} shows the unit cell integration of (a) charge density and (b) electron chirality for $\mrm{Te}$ as functions of chemical potential, which is denoted by the energy $\ep$.
To begin, we explain the results for charge density in (a) as a reference.
The total charge is the increasing function of $\ep$, and its derivative is the density of states, both of which provide useful information to understand the materials properties.

In analogy with $N$ and $\del N/\del \ep$, we calculate the total chirality $C$ and its energy derivative $\del C/\del \ep$, as shown in Fig.~\ref{fig:integral} (b).
The total chirality $C$ has non-zero values for the chiral crystal $\mathrm{Te}$.
Furthermore, $C$ has opposite signs with the same absolute value for the right- (green solid line) and left-handed (yellow dashed line) crystals.
This indicates that the total chirality $C$ serves as a measure of the handedness of the crystal structure.
Notably, the total chirality $C$ has rapid changes with respect to $\ep$, and the sign reverses.
Namely, even in a single chiral crystal, the sign of electron chirality is not uniquely determined and can reverse depending on the energy.

The energy derivative of the total chirality, $\del C/\del \ep$, is shown at the bottom of Fig.~\ref{fig:integral} (b).
The drastic changes in the total electron chirality $C$ occur at points where $\del C/\del \ep$ is large, as indicated by the black arrows for example.
Due to the significant variations in $C$ and $\del C/\del \epsilon$, the sign of the total chirality can be controlled using only a single right-handed (or left-handed) material, for instance, through electron or hole doping.

\paragraph{Photoemission spectroscopy.}

\begin{figure}[tb]
    \centering
    \includegraphics[width=8.5cm]{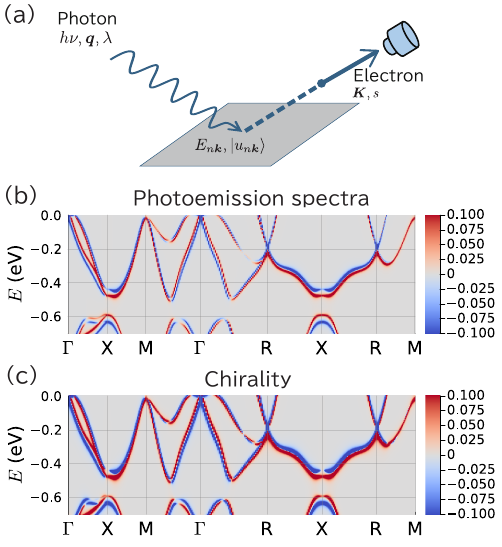}
    \caption{
    (a) Schematic illustration of photoemission spectroscopy with circularly polarized light.
    (b) Circular dichroism spectra for chiral crystal CoSi
    (c) Band dispersion with a characterization based on the electron chirality in $\bm k$-space for CoSi.
    We set the parameters as $\hat{\bm q} = \hat{\bm z}, \bm e_{\mrm L} = \hat{\bm x} + \imu\hat{\bm y}, \hbar\nu = 100 \, \mrm{eV}$ in (b).
    The magnitude of electron chirality in (c) is normalized by its maximum value in this energy range.
    }
    \label{fig:arpes}
\end{figure}

Phenomena unique to chiral materials can be captured through electron chirality-related observations. Consider, for example, photoemission spectroscopy.
The schematic figure is shown in Fig.~\ref{fig:arpes} (a).
The incident photon has energy $h\nu$, momentum $\bm q$, and polarization $\lam$, which is denoted by $\ket{\mrm i}$.
Then, the photon scatters with a $n$-th Bloch state electron with energy $E_{n\bm k}$ in the material.
Finally, the electron is emitted as a plane wave with momentum $\bm K$ and spin $s$, which is represented as the final state $\ket{f}$.
$\bm K$ satisfies $\bm K = \bm q + \bm k + \bm G$ ($\bm G$ is a reciprocal vector) due to the conservation of momentum.

The photoemission spectra for $\lam$-polarized light is proportional to the transition probability $I_{\lam} \propto \frac{2\pi}{\hbar} \sum_{\mrm i \mrm f} |\braket{\mrm f|\mscr H_{\mrm{ext}}|\mrm i}|^2 \delta(E_{\mrm f} - E_{\mrm i})$ \cite{Pendry76, Damascelli03, Sobota21}.
For electron-photon coupling Hamiltonian $\mscr H_{\mrm{ext}}$, we consider $\mscr H_{\mrm{ext}} = -\int \diff \bm r [\frac{1}{c}\bm j \cdot \bm A + \bm M_S \cdot \bm B]$ with $\bm j = \frac{e}{2m} \psi^\dg \dvec{\bm p} \psi$, $\bm M_S = \frac{\hbar e}{2mc} \psi^\dg \bm \sg \psi$, and $\bm B = \nabla \times \bm A$ \cite{Multunas23}, where the second term is a Zeeman term known as one of the relativistic effect.
We here consider the left-/right-circularly polarized light, and we represent it as $\lam = \mrm{L}/\mrm{R}$.
The circular dichroism photoemission spectrum is defined by the difference of the spectra for left- and right-polarized light $I_{\mrm{CD}} = I_{\mrm L} - I_{\mrm R}$.
We assume that we observe the photoemission electron with spin $s = \ua, \da$ states for all possible momenta $\bm K$, and then we obtain
\begin{align}
    I_{\mrm{CD}}^{\bm q}(E, \bm k) \propto -\sum_{n i j} \Big[j_{Sn\bm k ij}
    + \bm s_{n\bm k} \cdot \bm q \delta_{ij} \Big] \mcal E_{ij}^{\bm q} \nt
    \times f(E) \delta(E - E_{n\bm k}) \label{eq:arpes}
\end{align}
with $\mcal E_{ij}^{\bm q} = \imag[ e_{\mrm{R}\bm q, i} (\hat{\bm q} \times \bm e_{\mrm{L}\bm q})_j]$.
We have introduced Bloch function-based representations for the spin current tensor $j_{Sn\bm k ij} = \int\diff\bm r \psi_{n\bm k}^\ast \dvec{p_i} \sg^j \psi_{n\bm k}$ and the spin $\bm s_{n\bm k} = \frac{\hbar}{2} \int\diff\bm r\psi_{n\bm k}^\ast \bm \sg  \psi_{n\bm k}$.
The first term of Eq.~\eqref{eq:arpes} appears due to the cross term of $\bm j \cdot \bm A$ part and Zeeman term, namely relativistic effect.
The detailed derivation of Eq.~\eqref{eq:arpes} is explained in the SM \cite{supp}.
Note that $j_{Sn\bm kij}$ can be generally decomposed as $j_{Sn\bm kij} = \frac{1}{3}\tau_{n\bm k}^Z\delta_{ij} - \frac{1}{2} \sum_k \ep_{ijk} \mcal P_{Sn\bm k, k} + j_{Sn\bm kij}^{\mrm{sym}}$, where the first term is the electron chirality, the second term is the spin-derived electric polarization, and the third term is the residual symmetric part.
The matrix element $\mcal E_{ij}^{\bm q}$ represents the contribution of the photon, and is symmetric with respect to $i, j$.
Thus, $I_{\mrm{CD}}$ in Eq.~\eqref{eq:arpes} can capture the chirality $\tau_{n\bm k}^Z$ and symmetric part $j_{S\bm kij}^{\mrm{sym}}$.

Since the observed spectra depend on the angles of the momentum of incident photons, the first term in Eq.~\eqref{eq:arpes} includes the off-diagonal part with respect to $i, j = x, y, z$ in general. 
On the other hand, when we consider the isotropic spectrum $\sum_{\bm q = q\hat{\bm x}, q\hat{\bm y}, q\hat{\bm z}} I^{\bm q}_{CD}$, the first term of Eq.~\eqref{eq:arpes} becomes electron chirality because of the relation $\sum_{\bm q = q\hat{\bm x}, q\hat{\bm y}, q\hat{\bm z}} \mathcal E_{ij}^{\bm q} \propto \delta_{ij}$.

Figure.~\ref{fig:arpes} (b) shows the numerical result of the circular dichroism spectra $I_{\mrm{CD}}$, and Fig.~\ref{fig:arpes} (c) shows the band dispersion with a characterization based on the 
chirality $\tau_{n\bm k}^Z$ for chiral crystal $\mrm{CoSi}$.
The horizontal axes in both Figs.~\ref{fig:arpes} (b) and (c) are the path along the high symmetry points in the Brillouin zone.
For plot in Fig.~\ref{fig:arpes} (b), we set $\hat{\bm q} = \hat{\bm z}, \bm e_{\mrm L} = \hat{\bm x} + \imu\hat{\bm y}$, and obtain $\mcal E_{ij}^{\bm q} = \mrm{diag}(1, 1, 0)$.
These two figures have a similar structure for $\hbar\nu = \hbar|\bm q|c = 100\, \mrm{eV}$, because the second term in Eq.~\eqref{eq:arpes} is relatively small for $\mrm{CoSi}$.
In this way, electron chirality is reflected in the photoemission spectra.

In Fig.~\ref{fig:arpes} (c), the two bands split by SOC have positive and negative electron chirality, respectively.
Consequently, the chirality characterizes the band splitting induced by SOC.
This fact explains the drastic changes in the energy dependence of $C$ and $\del C/\del \ep$ shown in Fig.~\ref{fig:integral} (b) by noting the relation Eq.~\eqref{eq:chirality_bnd}.
Namely, when the chemical potential crosses a band with positive (negative) electron chirality, $\del C/\del \ep$ takes on a positive (negative) value.
Thus, $\del C/\del \ep$ can have a large change with respect to the chemical potential, when bands with positive and negative chirality are close in energy, as shown in Fig.~\ref{fig:arpes} (c).

Let us comment on the comparison with experimental observations.
In Ref.~\cite{Brinkman24}, photoemission spectra were measured for right- and left-handed circularly polarized light, and the difference between these spectra, namely the circular dichroism spectra, was discussed. 
This spectrum qualitatively corresponds to $I_{\mrm{CD}}$.
In both our calculation and the experimental results, the spectra for right- and left-handed crystals have opposite signs. 
This indicates the electron chirality $\tau^Z$ can be directly observed via the circular dichroism, although it is necessary to take into account surface effects and the work function for a detailed comparison, for example, based on the three-step model \cite{Damascelli03, Sobota21}.

\paragraph{Summary and discussion.}

We have studied chiral, axial, and polar crystals, proposing a quantitative characterization of these asymmetric materials based on the four-component Dirac field in relativistic quantum theory. 
The spatial distribution of electron chirality $\gamma^5$ is used to quantify electron chirality and electron axiality, in a manner similar to how charge density distribution characterizes polar crystals. 
Experimentally, electron chirality is linked to differences in photoelectric effects between right- and left-handed circularly polarized light. 
Additionally, using the polar vector defined by the Dirac field, we show that spin-derived electric polarization can quantify the polarity of polar crystals. 
This approach allows for quantifying asymmetry in materials, aiding the exploration of functionalities in chiral, axial, and polar crystals.

By systematically and quantitatively analyzing the distribution of electron chirality and the relation to response coefficients, various predictions concerning chiral properties will be made. While the cross-correlated response functions and transport coefficients can also serve as a quantitative measure of the asymmetric crystals, their magnitudes are influenced by multiple factors, including the density of states and relaxation time, in addition to the low-symmetry crystalline structure.
Hence, it is desirable to characterize a pure chirality which does not depend on extrinsic factors.
In this paper, we have proposed that the electron chirality and polarization, defined in terms of the Dirac field, serve this purpose.
Although we have focused on the time-reversal symmetric system in this study, the other types of microscopic quantities in relativistic quantum theory can be useful for characterizing time-reversal symmetry-broken systems.
A set of physical quantities is thoroughly summarized in a separate publication \cite{Hoshino24}.

\paragraph{Note added.}
After completion of our work, we became aware of the related work in Ref.~\cite{Sayantika24}, which studies the axial materials from first principles calculation.

\paragraph{Acknowledgement.}
The authors thank M. Senami and M. Fukuda for their useful comments.
TM is grateful to Y. Nomura and H.-Y. Chen for fruitful discussions.
This work was supported by the KAKENHI Grants No.~23KJ0298 (TM), No.~23K25827 (HI, MTS, SH), No.~24K00588 (MTS), No.~24K00581 (MTS), No.~24K00578 (SH).

\bibliography{dftchirality}
\bibliographystyle{apsrev4-2}

\clearpage
\appendix
\setcounter{equation}{0}
\setcounter{figure}{0}
\renewcommand{\theequation}{S\arabic{equation}}
\renewcommand{\thefigure}{S\arabic{figure}}

\noindent
{\large{\bf Supplementary Material for}} \\
\noindent
{\large{\bf ``Quantification of electronic asymmetry: chirality and axiality in solids''}}\\
\noindent
T. Miki, H. Ikeda, M.-T. Suzuki, and S. Hoshino

\section{A. Derivation of non-relativistic limit}

We derive the non-relativistic limit (NRL) representation for electron polarization \cite{Wang06}, and electron chirality \cite{Hoshino23, Hoshino24}.
Here, we explain the derivation in terms of left- and right-electron representation.

The expansion from non-relativistic limit for right-handed spinor field becomes $\psi_{\mrm R} \simeq (1 + \frac{\bm p\cdot \bm \sg}{2mc}) \psi + O(c^{-2})$ as derived from the Dirac equation.
Multiplying the three-component vector $\bm \sg$ (Pauli matrices), we obtain
\begin{align}
 \bm \sg \psi_{\mrm R}(\bm r) &= \frac{1}{\sqrt 2} \left( \bm \sg + \frac{\bm p}{2mc} + \frac{\imu}{2mc} \bm p \times \bm \sg \right) \psi(\bm r) 
 \label{eq:sigma_R_rep}
\end{align}
Namely, $\bm \sg \psi_{\mrm R}$ encloses the information of spin ($\bm \sg \psi$), momentum or particle current ($\bm p \psi$), and polarization ($\bm p\times \bm \sg\psi$).
The intuitive interpretation for $\bm p\times \bm \sg$ is gained in analogy to the Faraday effect~\cite{Hoshino23}, and the third term in Eq.~\eqref{eq:sigma_R_rep} gives a part of leading-order contribution for $\bm{\mcal P}$.
The representation for left-handed particles is also obtained in a similar manner.

\section{B. Details of first principles calculation}

We performed the density functional calculations using \texttt{Quantum ESPRESSO} (QE) \cite{Giannozzi17}.
In QE calculations, we used the exchange-correlation functional proposed by Perdew, Burke, and Ernzerhof \cite{Perdew96}, and optimized norm-conserving Vanderbilt pseudopotential \cite{Hamann13} provided in \texttt{PseudoDojo} \cite{vanSetten18}.

For $\mrm{Te}$, the plane wave cutoff energy was set to be $70\, \mrm{Ry}$, and the number of $\bm k$-points are taken as $8 \times 8 \times 8$.
We use modified lattice constants, which are obtained from Ref.~\cite{Nakazawa24}, to reproduce the band gap at $\mrm H$ point.
For other materials, we use the following plane wave cutoff energies and the number of $\bm k$-points: $80\, \mrm{Ry}$ cutoff energy, $6 \times 6 \times 4$ $\bm k$-points for $\mrm{K_2Zr(PO_4)_2}$, and $90\, \mrm{Ry}$ cutoff energy, $8 \times 8 \times 8$ $\bm k$-points for $\mrm{CoSi}$ and $\mrm{BaTiO3}$.

In the calculation of Fig.~3, Fig.~\ref{fig:polar} (b), and Fig.~\ref{fig:axial}, we use the Wannier interpolation.
The Wannier function is chosen as the closest to the closest 
to hydrogenic-atom orbitals \cite{Martin_book, Carlson57, Ozaki23}.
We choose the orbital as follows: Te ($3p$) for Te, O ($2p$) and Ti ($3d_{xy}, 3d_{yz}, 3d_{xz}$) for $\mathrm{BaTiO_3}$, and O ($2p$) for $\mathrm{K_2Zr(PO_4)_2}$.

We use the Methfessel-Paxton first-order spreading \cite{Methfessel89} for smearing in QE calculation, Fig.~3, Fig.~\ref{fig:polar} (b), and Fig.~\ref{fig:axial}.

\section{C. Spin-derived electric polarization in BaTiO${}_{\bf 3}$}

\begin{figure}[H]
    \centering
    \includegraphics[width=8.5cm]{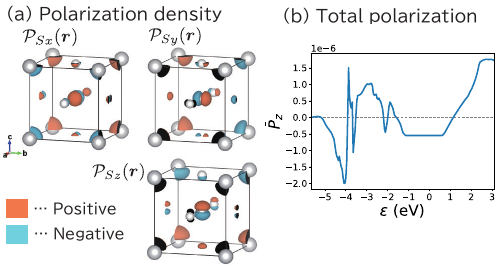}
    \caption{
    (a) Spin-derived electric polarization density
    (b) Energy dependences of total polarization for $\mrm{BaTiO_3}$.
    }
    \label{fig:polar}
\end{figure}

In this section, we focus on the electric polarization $\bm{\mcal P}(\bm r)$ defined in Eq.~\eqref{eq:polar} in the main text.
As mentioned in the main text, $\bm{\mcal P}(\bm r)$ is divided into the gradient of electronic density and the spin-derived electric polarization $\bm{\mcal P}_S(\bm r)$ in the NRL.

Figure~\ref{fig:polar} (a) shows the spatial distribution of $\bm{\mcal P}_S(\bm r)$.
$\mcal P_{Sx}(\bm r)$is polarized along the $x$-axis, with positive and negative values distributed in the $x$-direction.
Similarly, $\mcal P_{Sy}(\bm r)$ is polarized along the $y$-axis, and $\mcal P_{Sz}(\bm r)$ is polarized along the $z$-axis.

Since $\bm{\mcal P}_S(\bm r)$ does not have the dependence of unit cell origin, the spatial integral can be taken directly.
The chemical potential dependence of total polarization $\bar{\bm P} = \int_{\mrm{cell}} \diff\bm r \la \bm{\mcal P}_S(\bm r) \ra$ is shown in Fig.~\ref{fig:polar} (b).
Since the crystal structure of $\mrm{BaTiO_3}$ has rotational symmetry along the $z$-axis and then only $\bar{P}_{Sz}$ remains non-zero value, we show the $z$-component $\bar{P}_{Sz}$.

\section{D. Electron chirality for non-axial K${}_{\bf 2}$Zr(PO${}_{\bf 4}$)${}_{\bf 2}$}

\begin{figure}[H]
    \centering
    \includegraphics[width=8.5cm]{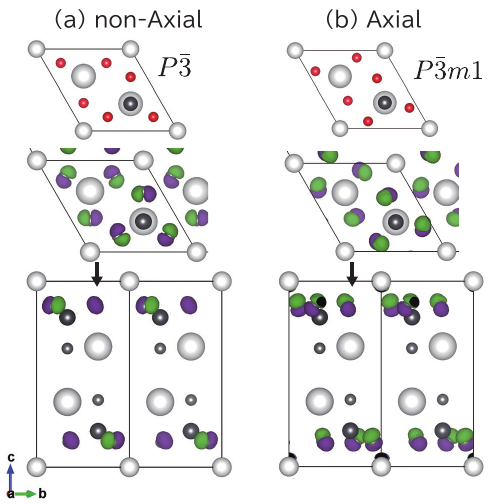}
    \caption{
    Spatial distributions of electron chirality $\tau^Z(\bm r)$ for $\mrm{K_2Zr(PO_4)_2}$ in (a) non-axial and (b) axial atomic configuration.
    The bottom figure of (b) is adopted from Fig.~\ref{fig:space} (b).
    }
    \label{fig:chirality}
\end{figure}

$\mrm{K_2Zr(PO_4)_2}$ exhibits two types of crystal structures: one with an axial atomic configuration ($P\bar 3$) and another with a non-axial atomic configuration ($P\bar 3 m1$).
The top figures in Figs.~\ref{fig:chirality} (a) and (b) represent the top view of the crystal structure.
The only difference between the two crystal structures is the positions of the oxygen atoms surrounding the phosphorus atoms, as indicated by the red points.

In the bottom figures of Fig.~\ref{fig:chirality}, we present the spatial distribution of $\tau^Z(\bm r)$ for $\mathrm{K_2Zr(PO_4)_2}$. 
Figure~\ref{fig:chirality} (a) shows $\tau^Z(\bm r)$ for a non-axial atomic configuration, and Fig.~\ref{fig:chirality} (b) provides it for the axial configuration as a reference.
In both non-axial and axial configurations, the spatial distribution of chirality $\tau^Z(\bm r)$ indicates the presence of chirality dipoles.
In the non-axial configuration in Fig.~\ref{fig:chirality} (a), however, the dipole is tilted.

\section{E. Axiality in K${}_{\bf 2}$Zr(PO${}_{\bf 4}$)${}_{\bf 2}$}

\begin{figure}[ht]
    \centering
    \includegraphics[width=8.5cm]{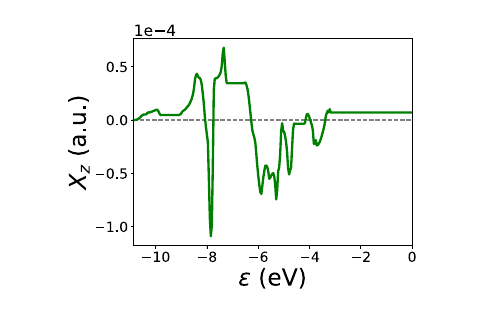}
    \caption{
    Energy dependences of total axiality for $\mrm{K_2Zr(PO_4)_2}$.
    }
    \label{fig:axial}
\end{figure}

Here, we consider the total axiality defined by $\bm X = \int_{\mrm{cell}} \diff\bm r \la \bm r \tau^Z(\bm r) \ra$.
Figure~\ref{fig:axial} shows the chemical potential dependence of $z$-component of the axiality.
For the plot, we set the spatial origin at the $\mrm{Zr}$ cite, which is one of the inversion centers, and the unit cell is chosen as the Wigner-Seitz cell.
In this choice, the $x,y$-component of $\bm X$ becomes zero because of $3$-fold rotational symmetry along $z$-direction.
On the other hand, the $z$-component has non-zero value for the axial crystal $\mrm{K_2Zr(PO_4)_2}$.
We note that if we fix the unit cell in the inversion-symmetric system, the value of $\bm X$ does not depend on the position of origin because of the relation $\int_{\mrm{cell}} \diff\bm r \la(\bm r - \bm r_0) \tau^Z(\bm r)\ra = \bm X - \bm r_0C$ and $C = 0$. 
This property is unique to axiality and does not appear in the polarity $\bm P$.

\section{F. Derivation of photoemission spectra}

In this section, we explain the detailed derivation of Eq.~\eqref{eq:arpes}.
As mentioned in the main text, we evaluate the quantity \cite{Pendry76, Damascelli03, Sobota21}:
\begin{align}
    \sum_{\mrm i \mrm f} \frac{2\pi}{\hbar} |\braket{\mrm f | \mscr H_{\mrm{ext}} | \mrm i}|^2 \delta(E_{\mrm f} - E_{\mrm i}) \label{eq:fermi_golden_rule}
\end{align}
Here, we reconsider the electron-photon coupling Hamiltonian \cite{Multunas23} given by
\begin{align}
    \mscr H_{\mrm{ext}} = -\int \diff \bm r \bigg[\frac{1}{c}\bm j \cdot \bm A + \bm M_S \cdot \bm B\bigg] \label{eq:hext}.
\end{align}
The vector potential $\bm A$ is composed of two types of polarized photons:
\begin{align}
    \bm A(\bm r) = \sqrt{4\pi c} \int \frac{\diff\bm q}{(2\pi)^3} \sqrt{\frac{\hbar}{2\om_{\bm q}}} \epn^{\imu\bm q \cdot \bm r} \sum_{\lam} \bm e_{\lam \bm q} a_{\lam \bm q} + \Hc
\end{align}
where $a_{\lam \bm q}$ is a annihilation operator of photon, and $\om_{\bm q} = c|\bm q|$ is a dispersion of photon.

We set the initial and final states as $\ket{\mrm i} = \ket{\mrm{GS}} a_{\lam \bm q}^\dg \ket{0}$ and $\ket{\mrm f} = c_{\mrm{out} \bm K, s}^\dg c_{n \bm k} \ket{\mrm{GS}} \ket{0}$, respectively, where $\ket{\mrm{GS}}$ indicates electronic ground state.
$\bm K$ can be decomposed by $\bm K = \bm k' + \bm G + \bm q$, where $\bm G$ is a reciprocal vector and $\bm k'$ is a wave vector inside the Brillouin zone, without loss of generality.
We expand Bloch function by the plane wave as $\psi_{n\bm k}(\bm r s) = (1/\sqrt V)\sum_{\bm G} C_{n\bm k}^s(\bm G) \epn^{\imu(\bm k + \bm G) \cdot \bm r}$.
Then, $\braket{\mrm f| \mscr H_{\mrm{ext}} |\mrm i}$ in Eq.~\eqref{eq:fermi_golden_rule} reduces to
\begin{align}
    \braket{\mrm f| \mscr H_{\mrm{ext}} |\mrm i}
    &\propto \sqrt{\frac{1}{\om}} \sum_{s_1} C_{n\bm k}^{s_1}(\bm G) [(\bm k + \bm G) \cdot \bm e_{\lam \bm q} \delta_{ss_1} \nt
    &+ \frac{1}{2}\bm \sg_{s s_1} \cdot (\imu\bm q \times \bm e_{\lam \bm q})] f_{n\bm k} \delta_{\bm k \bm k'} \label{eq:fhi}
\end{align}
The spectra Eq.~\eqref{eq:arpes} in the main text can be calculated by inserting Eq.~\eqref{eq:fhi} into Eq.~\eqref{eq:fermi_golden_rule} and decomposing by energies.

\end{document}